\documentclass[
    ,final            
  ]
  {aipproc}

\layoutstyle{6x9}

\begin{document}

\title{Extracting SUSY parameters from LHC measurements using Fittino}


\author{Philip Bechtle}{
  address={DESY, Notkestr.~85, 22607 Hamburg, Germany}
}

\author{Klaus Desch}{
  address={Dept.~of Physics, University of Bonn, Nussallee 12, 53115
Bonn, Germany} }

\author{Mathias Uhlenbrock}{
  address={Dept.~of Physics, University of Bonn, Nussallee 12, 53115
Bonn, Germany} }

\author{Peter Wienemann}{
  address={Dept.~of Physics, University of Bonn, Nussallee 12, 53115
Bonn, Germany} }

\begin{abstract}
We show that presently available precision data are in good agreement
with supersymmetry at a mass scale below 1~TeV. Using a SUSY point
close to the best fit to present data, we give a projection of the
capabilities of the LHC to constrain SUSY models and their parameters
as function of the accumulated luminosity.
\end{abstract}

\maketitle


\section{Introduction}

With the upcoming start of the Large Hadron Collider (LHC) we will be
in the exciting situation to be able to probe physics at the TeV
energy scale for the first time directly in a laboratory. There are
numerous, very diverse reasons to expect new physics showing up at
this energy scale. One of the most promising candidates for physics
beyond the standard model is supersymmetry (SUSY). It is able to
remedy various different shortcomings of the standard model by
introducing a single additional symmetry between bosons and
fermions. 

If new physics can be established at the LHC, the most important
objective will be to find out the underlying model and to determine
its theory parameters. For such an inverse modelling task fitting
packages like {\tt Fittino}~\cite{Bechtle:2004pc,Bechtle:2005vt} and
{\tt SFitter}~\cite{Lafaye:2007vs} have been developed. For the
results described in this work, {\tt Fittino} has been
employed. Sparticle properties for given SUSY parameters are
calculated by the programs {\tt SPheno}~\cite{Porod:2003um} and {\tt
  ``Mastercode''}~\cite{Buchmueller:2007zk,Buchmueller:2008qe,Buchmueller:2009fn}. Starting
from a model and parameters which show good agreement with available
measurements from LEP, SLC, Tevatron, $B$ factories, WMAP and
precision measurements at low energy, we present a projection of the
expected inverse modelling performance as function of the accumulated
LHC luminosity. A detailed description of this work can be found
in~\cite{Bechtle:2009ty}.

\section{Fits to low-energy data}

Although so far there is no unambiguous experimental evidence for
supersymmetric particles, there are a number of measurements which
allow to put constraints on a supersymmetric theory. These include
precision measurements and Higgs boson mass limits from the high
energy colliders LEP, SLC and Tevatron, as well as information on rare
decays of $K$ and $B$ mesons, the anomalous magnetic moment of the
muon and the cold dark matter relic density as derived from
cosmological measurements. A complete list of used measurements can be
found in~\cite{Bechtle:2009ty}.

Various SUSY models are fitted to these measurements and constaints on
the respective model parameters are derived. The tested models
comprise mSUGRA with sign$(\mu) = +1$, mSUGRA with sign$(\mu) = -1$,
GMSB with sign$(\mu) = +1$ and $N_5 = 1,2,3,4$ and GMSB with
sign$(\mu) = -1$ and $N_5 = 1$. Since the lightest SUSY particle (LSP)
in GMSB models is a very light gravitino making up hot instead of cold
dark matter, the cold dark matter relic density measurement is
excluded from the list of observables for GMSB fits. It turns out that
mSUGRA with sign$(\mu) = +1$ is in good agreement with the data. The
corresponding best fit parameters are listed in
Table~\ref{tab:LEFitResult}. The sparticle masses corresponding to
these parameters are all predicted to be below 1~TeV and should thus
be discoverable at the LHC rather early.
\begin{table}
\begin{tabular}{l r c l r} \hline
Parameter & Best fit & & Uncertainty & SPS1a\\ \hline
$\tan \beta$ & 13.2 & $\pm$ & 7.2 & 10\\
$M_{1/2}$ (GeV) & 331.5 & $\pm$ & 86.6 & 250 \\
$M_0$ (GeV) & 76.2 & $\pm$ & $^{+79.2}_{-29.1}$ & 100 \\
$A_0$ (GeV) & 383.8 & $\pm$ & 647 & $-100$ \\ \hline
\end{tabular}
\caption{Fitted mSUGRA parameters assuming sign$(\mu) = +1$ together
  with SPS1a values.}
\label{tab:LEFitResult}
\end{table}

It is striking that the fitted parameter point is in rather good
agreement with the well studied SPS1a benchmark
scenario~\cite{Allanach:2002nj}. For this benchmark point there is a
wealth of detailed experimental Monte Carlo studies available. Apart
from a significantly larger cold dark matter relic density, the
phenomenology of SPS1a is very similar to the best fit point. We
therefore use the expected measurements for SPS1a from the available
experimental studies in order to try a projection of the low-energy
fit results to the LHC era.

\section{Expected constraints from LHC data}

The projections are performed for three different integrated LHC
luminosities: 1~fb$^{-1}$, 10~fb$^{-1}$ and 300~fb$^{-1}$. All results
assume a centre-of-mass energy of 14~TeV. Wherever possible directly
measurable observables are used as input to the fit. Many of them are
endpoints of mass spectra. Most often the decay chain $\tilde{q}_L \to
\tilde{\chi_2^0} q \to \tilde{\ell}^{\pm}_R \ell^{\mp} q \to
\tilde{\chi}^0_1 \ell^+ \ell^- q$ is exploited. Also mass peaks and
two ratios of branching fractions are used.  A full list of used input
observables can again be obtained from~\cite{Bechtle:2009ty}.

\subsection{mSUGRA model fits}

The expected parameter uncertainties for an mSUGRA fit with fixed
sign$(\mu)$ is shown in Table~\ref{tab:mSUGRAFitResult}. The scalar
mass parameter $M_0$ and the gaugino mass parameter $M_{1/2}$ can
already be constrained to the level of a few percent with an
integrated luminosity of 1~fb$^{-1}$. More challenging are $\tan
\beta$ and $A_0$. The uncertainties go down with increasing
luminosity. At 300~fb$^{-1}$, $M_0$ and $M_{1/2}$ finally reach the
(few) permille level and for $\tan \beta$ ($A_0$) a relative precision
of 4~\% (11~\%) is achieved.
\begin{table}
\begin{tabular}{l r r r r} \hline
          &       &\multicolumn{3}{c}{Uncertainties}  \\ 
Parameter & SPS1a & 1~fb$^{-1}$ & 10~fb$^{-1}$ & 300~fb$^{-1}$ \\ \hline
sign$(\mu)$ & $+1$ & & & \\
$\tan \beta$ & 10 & 3.7 & 0.84 & 0.35 \\
$M_{1/2}$ (GeV) & 250 & 6.7 & 1.2 & 0.30 \\
$M_0$ (GeV) & 100 & 4.2 & 2.1 & 0.39 \\
$A_0$ (GeV) & $-100$ & 742.1 & 52.9 & 11.1 \\ \hline
\end{tabular}
\caption{Expected uncertainties on mSUGRA parameters for integrated
  LHC luminosities of 1~fb$^{-1}$, 10~fb$^{-1}$ and 300~fb$^{-1}$.}
\label{tab:mSUGRAFitResult}
\end{table}

To determine sign$(\mu)$, two fits are performed to each (of many) toy
datasets obtained by smearing observable values around the values for
the best fit point. One fit assumes sign$(\mu) = +1$ and the other one
uses sign$(\mu) = -1$.  The $\chi^2$ correlations of the fits with
sign$(\mu) = +1$ and sign$(\mu) = -1$ allow to choose the most
probable sign$(\mu)$ and to determine the probability for this choice
to be right or wrong. Already at 1~fb$^{-1}$ there is a good chance to
determine sign$(\mu)$ correctly. The probability for an incorrect
decision is less than 5~\%. For 10~fb$^{-1}$ or more luminosity,
sign$(\mu)$ can be determined with negligible error probability.

\subsection{MSSM18 model fits}

Performing an mSUGRA fit implies strong constraints on the sparticle
masses and couplings due to the assumed SUSY breaking mechanism. It is
also interesting to fit more general models to the data without making
assumptions on the high-scale behaviour and derive properties of SUSY
breaking from weak-scale parameters using a bottom-up approach. The
Lagrangian of the most general minimal supersymmetric standard model
(MSSM) introduces more than 100 SUSY parameters, but excluding
flavour-non-diagonal and CP-violating terms and assuming (effective)
universality of the first two generations reduces the number
of parameters to 18. We refer to this model as MSSM18.

\begin{table}
\begin{tabular}{lrcl}
  \hline
  Parameter & Nominal value & & Uncertainty (LE$+$LHC300)\\

  \hline
  $M_{\tilde{\ell}_L}$ (GeV)    &     194.31 & $\pm$ & 6.4 \\
  $M_{\tilde{\ell}_R}$ (GeV)    &     135.76 & $\pm$ & 10.5 \\
  $M_{\tilde{\tau}_L} $ (GeV) &   193.52 &  $\pm$ & 43.0 \\
  $M_{\tilde{\tau}_R} $ (GeV) &   133.43 &  $\pm$ & 38.2 \\
  $M_{\tilde{q}_L}    $ (GeV) &   527.57 & $\pm$ & 3.4 \\
  $M_{\tilde{q}_R}    $ (GeV) &     509.14 & $\pm$ & 9.0 \\
  $M_{\tilde{b}_R}    $ (GeV) &     504.01 & $\pm$ & 33.3 \\
  $M_{\tilde{t}_L}    $ (GeV) &     481.69 & $\pm$ & 15.5 \\
  $M_{\tilde{t}_R}    $ (GeV) &     409.12 & $\pm$ & 103.8 \\
  $\tan\beta          $ &    10      &  $\pm$ & 3.3\\
  $\mu                $ (GeV) &     355.05 &  $\pm$ & 6.2 \\
  $X_{\tau}           $ (GeV) & $-$3799.88   & $\pm$ & 3053.5 \\
  $X_{t}              $ (GeV) &  $-$526.62   & $\pm$ & 299.2 \\
  $X_{b}              $ (GeV) & $-$4314.33   & $\pm$ & 5393.6 \\
  $M_1                $ (GeV) &     103.15 &  $\pm$ & 3.5 \\
  $M_2                $ (GeV) &     192.95 &  $\pm$ & 5.5 \\
  $M_3                $ (GeV) &     568.87 &  $\pm$ & 6.9 \\
  $m_{A}            $ (GeV) &   359.63   & $\pm$ & $_{-99.3}^{+1181}$\\
  \hline
\end{tabular}
\caption{Expected uncertainties on MSSM18 parameters from a
  combination of low-energy and expected LHC measurements for an
  integrated luminosity of 300~fb$^{-1}$.}
\label{tab:MSSM18Results}
\end{table}

The outcome of an MSSM18 fit to combined low-energy and LHC data with
300~fb$^{-1}$ of luminosity is summarised in
Table~\ref{tab:MSSM18Results}. Some parameters are determined with
reasonable precision but there are also some which are only weakly
constrained. Particularly difficult to measure are those parameters
which characterise third generation sfermion properties and the Higgs
sector parameters. The reason for the weak constraints in these
sectors is that they are only partially accessible at the LHC. For the
considered parameter point, the heavy Higgs bosons, for instance, are
not expected to be discovered at the LHC. The situation can be
improved significantly by including measurements at a future $e^+ e^-$
linear collider. In that case the precision on many parameters is one
or two orders of magnitude better.

\section{Conclusions}

Presently available precision data are in good agreement with SUSY
masses below 1~TeV and thus an early discovery at the LHC is
possible. For a SUSY point close to the best fit to present data, we
show that even in rather general models like the MSSM18, the LHC (with
ultimate luminosity) will allow for a determination of the SUSY
parameters with reasonable precision except for third generation
sfermion and Higgs sector parameters. In more constrained models like
mSUGRA the achievable precision will be typically higher by one order
of magnitude.


%



\bibliographystyle{aipproc}   




\end{document}